\documentclass{elsart}

\usepackage{graphicx,amssymb}
\usepackage[latin1]{inputenc}

\begin{document}

\begin{frontmatter}

\title{Ermakov Systems with Multiplicative Noise}

\author[label1]{E. Cervantes-L\'opez}\ead{cer.ern@gmail.com},
\author[label1]{P.B. Espinoza}\ead{wolfgang@culagos.udg.mx},
\author[label1]{A. Gallegos}\ead{gallegos@culagos.udg.mx},
\author[label2]{H.C. Rosu}\ead{hcr@ipicyt.edu.mx}

\address[label1]{Departamento de Ciencias Exactas y Tecnolog\'{\i}a,\ Centro Universitario de los Lagos, Universidad de Guadalajara, Enrique D\'{\i}az de Le\'{o}n 1144, Col. Paseos de la Monta\~{n}a, Lagos de Moreno, Jalisco, M\'{e}xico}
\address[label2]{Division of Advanced Materials, IPICyT, 78231 San Luis Potos\'{\i}, S.L.P., Mexico}

\begin{abstract}

 Using the Euler-Maruyama numerical method, we present calculations of the Ermakov-Lewis invariant and the dynamic, geometric, and total  phases for several cases of stochastic parametric oscillators, including the simplest case of the stochastic harmonic oscillator. The results are compared with the corresponding numerical noiseless cases to evaluate the effect of the noise. Besides, the noiseless cases are analytic and their analytic solutions are briefly presented. The Ermakov-Lewis invariant is not affected by the multiplicative noise in the three particular examples presented in this work, whereas there is a shift effect in the case of the phases.

\end{abstract}


\bigskip

{\bf Physica A 401 (2014) 141-147}

\medskip

\begin{itemize}

\item  {\small Multiplicative noise is introduced in the Ermakov systems of equations.}

\item  {\small The noise is added in the Euler-Maruyama numerical scheme.}

\item  {\small No effect found in the Ermakov invariant only when noise is put in both equations.}

\item  {\small Lewis-Riesenfeld phases show a small shift effect for reasonable noise amplitudes.}

\end{itemize}

\end{frontmatter}

\section{Introduction}

 Noisy harmonic oscillators are widely used simple models with many applications in physics, chemistry, biology, medicine, economics and sociology and have been recently reviewed in the book of Gitterman \cite{G-book}. On the other hand, it is well established that integrals of motion and dynamical invariants are related to the symmetries of the conservative dynamical systems and many of them have well-established phenomenological meaning. In particular, an invariant quantity introduced by Ermakov \cite{Er80} a long time ago and reobtained by Lewis \cite{Lewis1} by another method and in a different context has become a standard concept in the dynamical analysis of the important class of parametric oscillator systems. These systems are widespread in many areas of physics such as semiclassical theory of radiation, mechanical oscillations with time-dependent parameters, motion of charged particles in certain types of magnetic fields, cosmological models, and so on (see \cite{Er90} and references therein). In general, such kinds of systems are described by the following Newton type equation of motion
\begin{equation}\label{e1}
\ddot x+\Omega^2(t)x=0~,  
\end{equation}
where $\Omega(t)$ is the time-dependent frequency.
One can write the solution of (\ref{e1}) in the Milne form \cite{Milne}
\begin{equation}\label{e2}
x(t)=C\rho(t)\sin \left(\Theta_T(t)+\phi\right)~.
\end{equation}
$C$ and $\phi$ are arbitrary constants and the phase $\Theta_T(t)$ is given by
\begin{equation}\label{e3}
\Theta_T(t)= \int^t\frac{1}{\rho^2(t')}dt'
\end{equation}
and $\rho(t)$ is a solution of the Milne-Pinney equation
\begin{equation}\label{INT2}
\ddot {\rho}+\Omega^2(t)\rho=\frac {k}{\rho^3}~,
\end{equation}
where $k$ is an arbitrary real constant. It is also known how to express the function $\rho$ in terms of two linearly independent solutions of the $x$ oscillator \cite{Er90}.

For parametric oscillators governed by Eqs.~(\ref{e1}) and (\ref{INT2}) forming a so-called Ermakov system, one can introduce the Ermakov-Lewis invariant \cite{Er80,Lewis1} given by the following formula \cite{Eliezer} (henceforth we use $k=1$)
\begin{equation}\label{INT1}
I=\frac{1}{2}\left(\rho \dot{x} - \dot{\rho}x \right)^{2} + \frac{1}{2}\left(\frac{x}{\rho}\right)^{2}~.
\end{equation}

\noindent Moreover, the total phase $\Theta_T$ is the sum of a pure dynamical phase and a geometric phase. The latter is
a rather common concept for time-dependent systems ever since it has been introduced by Berry in 1984 in a quantum-mechanical context \cite{BE}. In the case of classical mechanical systems these geometric phases are also called Hannay angles \cite{H85}. The dynamical phase is related to the dynamical nature of the evolution of the system, while the geometric phase depends on the geometry and topology of the phase space trajectory as a function of the variation of the parameters of the system.
The dynamic and geometric phases \cite{Er90,E2,C4} are given by the following equations (henceforth we use $k=1$)
\begin{equation}\label{AND}
\Delta\Theta_{d}(t)=\int^{t}\left[\frac{1}{\rho^{2}}-\frac{1}{2}\frac{d}%
{dt^{\prime}}\left(\dot{\rho}\rho\right)
+\dot{\rho}^{2}\right]  dt^{\prime}
\end{equation}
and
\begin{equation}\label{ANG}
\Delta\Theta_{g}(t)=\int^t\left[\frac{1}{2}\frac{d}{dt^{\prime}}\left(\dot%
{\rho}\rho\right)-\dot{\rho}^{2}\right]dt^{\prime}~.
\end{equation}
By examining the last two formulas, one can notice that their sum is indeed the total phase. Notice also that for a constant $\rho$ the geometric phase is naught.

\bigskip

\section{Stochastic calculus}\label{Smat}

Although there is a huge literature on the Ermakov systems there are only a couple of papers on the Ermakov approach for stochastic oscillators. In the 1980's, Nassar introduced an Ermakov-Nelson stochastic process in the hydrodynamic interpretation of the Schroedinger equation \cite{N85}, while in 2005 Haas \cite{haas05} used the Haba-Kleinert averaging method \cite{HK02} for the stochastic quantization of time-dependent systems.

\medskip

Motivated by this scarcity and also by the natural question of how robust are the Ermakov quantities of the parametric oscillator (\ref{e1}) to stochastic noises, we present here a study of the effects of the multiplicative noise on the Ermakov quantities.
The noise will be taken as a Brownian random walk noise in the framework of the theory of stochastic processes \cite{E4}. This theory has been very supportive in modeling phenomena that do not exactly follow a continuous path and have small disturbances when they evolve in noise-perturbative conditions.

\medskip

\noindent For the numerical solution scheme, we use the Euler-Maruyama method \cite{E3,C1,C2}, in which a stochastic differential equation has the following form
\begin{equation}\label{EDE1}
dY_{t} = a(t,Y_{t})dt + b(t,Y_{t})dB_{t}~,
\end{equation}
where $B_{t}$ is the stochastic variable.
\noindent The numerical method to solve (\ref{EDE1}) is given by the following expression
\begin{equation}\label{EDE4}
Y_{n+1} = Y_{n} + a(t_{n},Y_{n})\Delta t_{n} + b(t_{n},Y_{n})\Delta B_{n}~,
\end{equation}
where $\Delta B_{n}=B_{t_{n+1}}-B_{t_{n}}$. For computational purposes it is useful to consider a discretized Brownian motion. We divide the time interval $[0, T]$ into $N$ equal subintervals by setting $\Delta t_{n}=t_{n+1}-t_{n}= T/N$ or $t_{n} = n \frac{T}{N}, n =0, . . . ,N$.

\noindent Further, due to the properties of Brownian motion we can simulate its values at the selected points by $B_{t_{n+1}} = B_{t_{n}} +\Delta B_{n}$, with $B_{t_{0}}=0$ and $\Delta B_{n}=\sqrt{\Delta t_{n}}Z_{n}$ where $Z_{n}$ is an independent random variable with normal distribution. The assumptions, convergence criteria and stability conditions are all fulfilled according to textbooks \cite{C2,C3}. The Euler-Maruyama method is widely used in stochastic mathematics but only occasionally one can find more applied papers. Recently, Wan and Yin used it in a study of the effects of Gaussian colored noise and noise delay on a nonlinear calcium oscillation system \cite{Wan-Yin}.\\

For the case of the parametric oscillators as given by equations (\ref{e1}) and (\ref{INT2}), we consider first their matrix formulation and then the corresponding stochastic variables and coefficients are identified in the following explicit forms
\begin{equation}\label{EDE5}
dX_{t}=\left(
\begin{array}
[c]{c}%
dx\\
d\dot{x}%
\end{array}
\right)  ,\ \ \ \ a\left(  t,X_{t}\right)  =\left(
\begin{array}
[c]{c}%
\dot{x}\\
-\Omega^{2}\left(  t\right)  x
\end{array}
\right)  ,\ \ \ \ b\left(  t,X_{t}\right)  =\left(
\begin{array}
[c]{c}%
0\\
-\alpha_{\Omega}x^m %
\end{array}
\right)%
\end{equation}
and
\begin{equation}\label{EDE6}
d\rho_{t}=\left(
\begin{array}
[c]{c}%
d\rho\\
d\dot{\rho}%
\end{array}
\right)  ,\ \ \ \ a\left(t,\rho_{t}\right)  =\left(
\begin{array}
[c]{c}%
\dot{\rho}\\
-\Omega^{2}\left(  t\right) \rho+\frac{1}
{\rho^{3}}%
\end{array}
\right)  ,\ \ \ \ b\left(t,\rho_{t}\right)  =\left(
\begin{array}
[c]{c}%
0\\
-\alpha_{\Omega}\rho^m%
\end{array}
\right)~,%
\end{equation}
\noindent where $\alpha_{\Omega}$ is the amplitude of the frequency noise \cite{E3-b}.
The parameter $m$ takes the value $0$ for the additive noise and can be any positive integer number bigger than the unity for the multiplicative cases. Of course, one can use any function $g(x)$ and $g(\rho)$ instead of $x^m$ and $\rho^m$ to study more general couplings of the noise \cite{G2}. Unfortunately, it is easy to see that the additive noise case leads to parametric equations of the form $\ddot x+\Omega^2(t)x=\xi$ and
$\ddot {\rho}+\Omega^2(t)\rho=\frac {k}{\rho^3}+\xi$ for which the expression of the Ermakov-Lewis invariant is more complicated \cite{takayama}. This is due to the fact that the noise term occurs as a true forcing term and not as a contribution to the time-dependent frequency parameter.
Moreover, the calculations for the formulas of the dynamic and geometric phases involve auxiliary equations related to the forcing term. Thus this case is left for a future investigation. Besides, for $m>1$ the oscillators have amplitude-dependent frequencies and therefore although they could be still called parametric they are nonlinear in the frequency too and the Ermakov-Lewis approach in this case is still under development. Therefore only the $m=1$ multiplicative case will be considered in the applications to follow because for this case the standard Ermakov-Lewis analysis holds. For this case, the noise appears as a random effect in time on the time-dependent frequency.

In addition, as one can see from (\ref{EDE5}) and (\ref{EDE6}), we use the same noise term in the two oscillators of the Ermakov system. In our numerical calculations, we also studied cases containing the noise either only in the linear oscillator (\ref{EDE5}) or the nonlinear one (\ref{EDE6}). For these cases, independently of the used seed, we have seen a stronger effect of the noise on the Ermakov-Lewis invariant. For illustration purposes, this feature is presented in the first application below but similar plots have been obtained in the other two applications. Besides, we have seen the same shifting effect on the phases in the presence of noise in the $\rho$ oscillator and no effect at all when the noise was in the linear oscillator, as expected because the phases depend only on the $\rho$ function. On the other hand, we noticed an interesting compensating effect of the noise terms in the Ermakov-Lewis invariant when the same noise is added in both oscillators.

\medskip

\section{Applications}

 We move now to three explicit applications for which we follow Eliezer and Gray \cite{Eliezer} and use the initial conditions $x(0)=1$, $\dot{x}(0)=0$ for the solution of (\ref{e1}), which imply similar initial conditions for the Milne-Pinney solution: $\rho(0)=1$ and $\dot{\rho}(0)=0$.

\subsection{Harmonic oscillator}

\begin{figure} [x!] 
\begin{center}
\resizebox*{0.30\textheight}{!}{
{\includegraphics{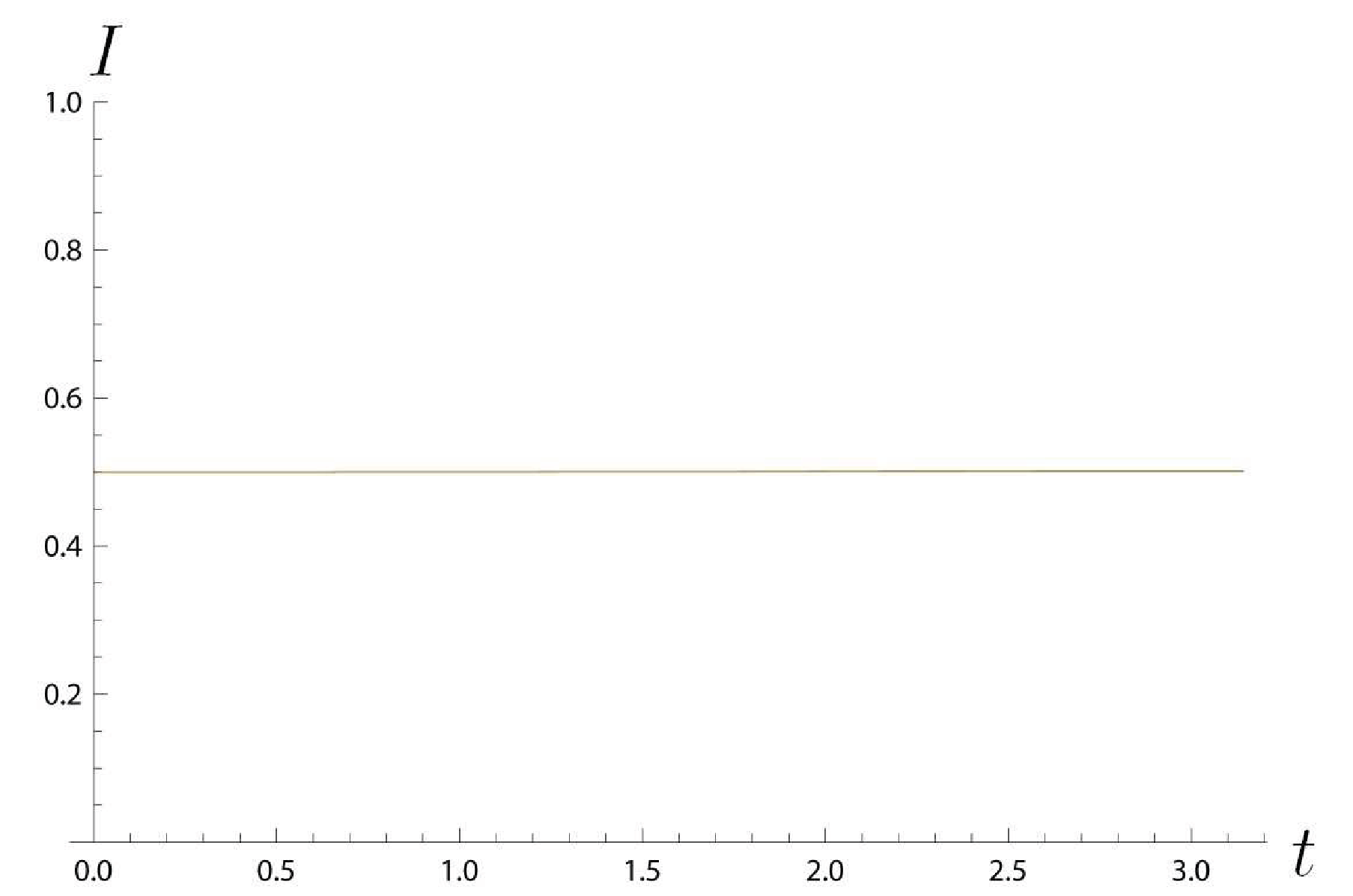}}}
\resizebox*{0.30\textheight}{!}{
{\includegraphics{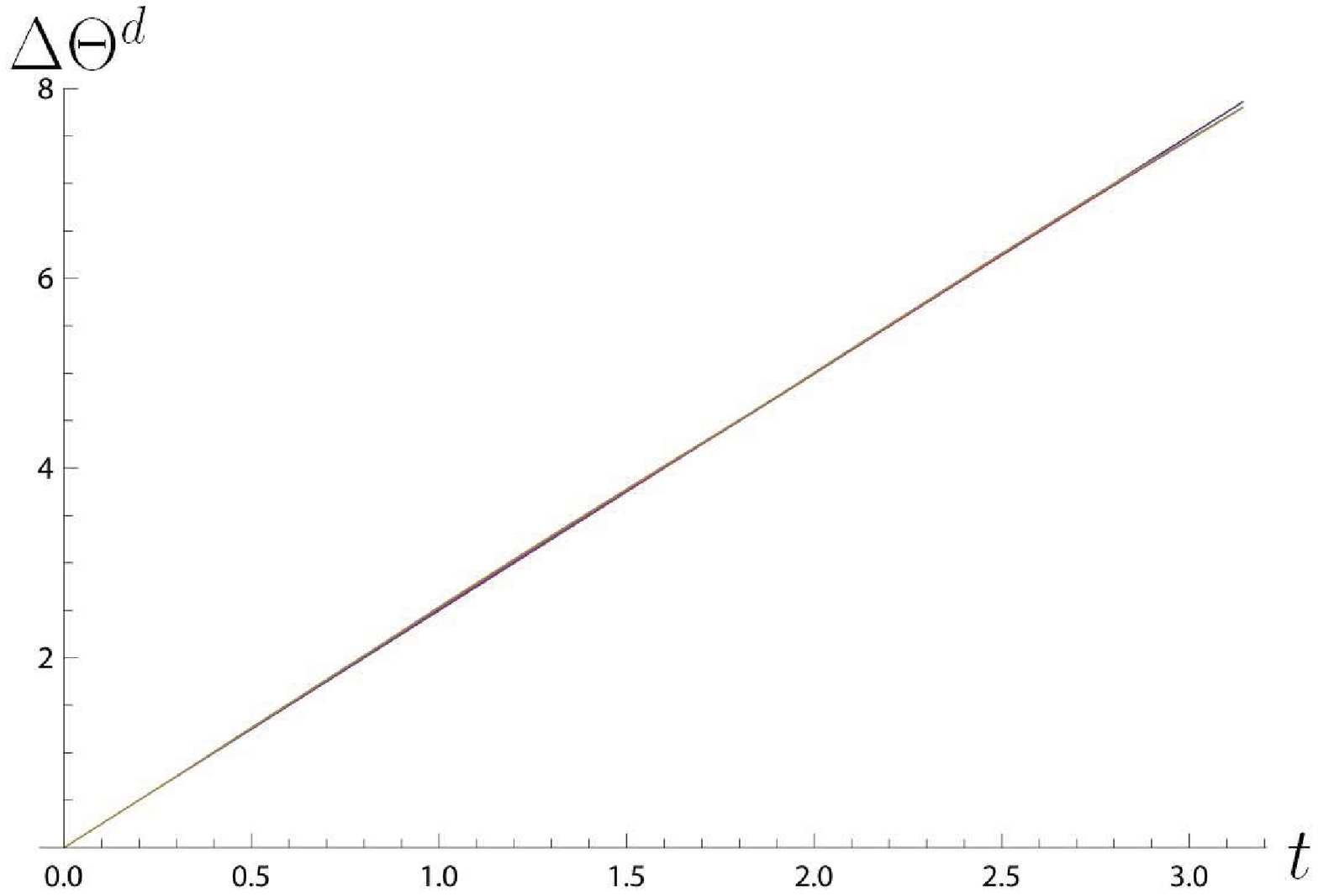}}}
\resizebox*{0.23\textheight}{!}{
{\includegraphics{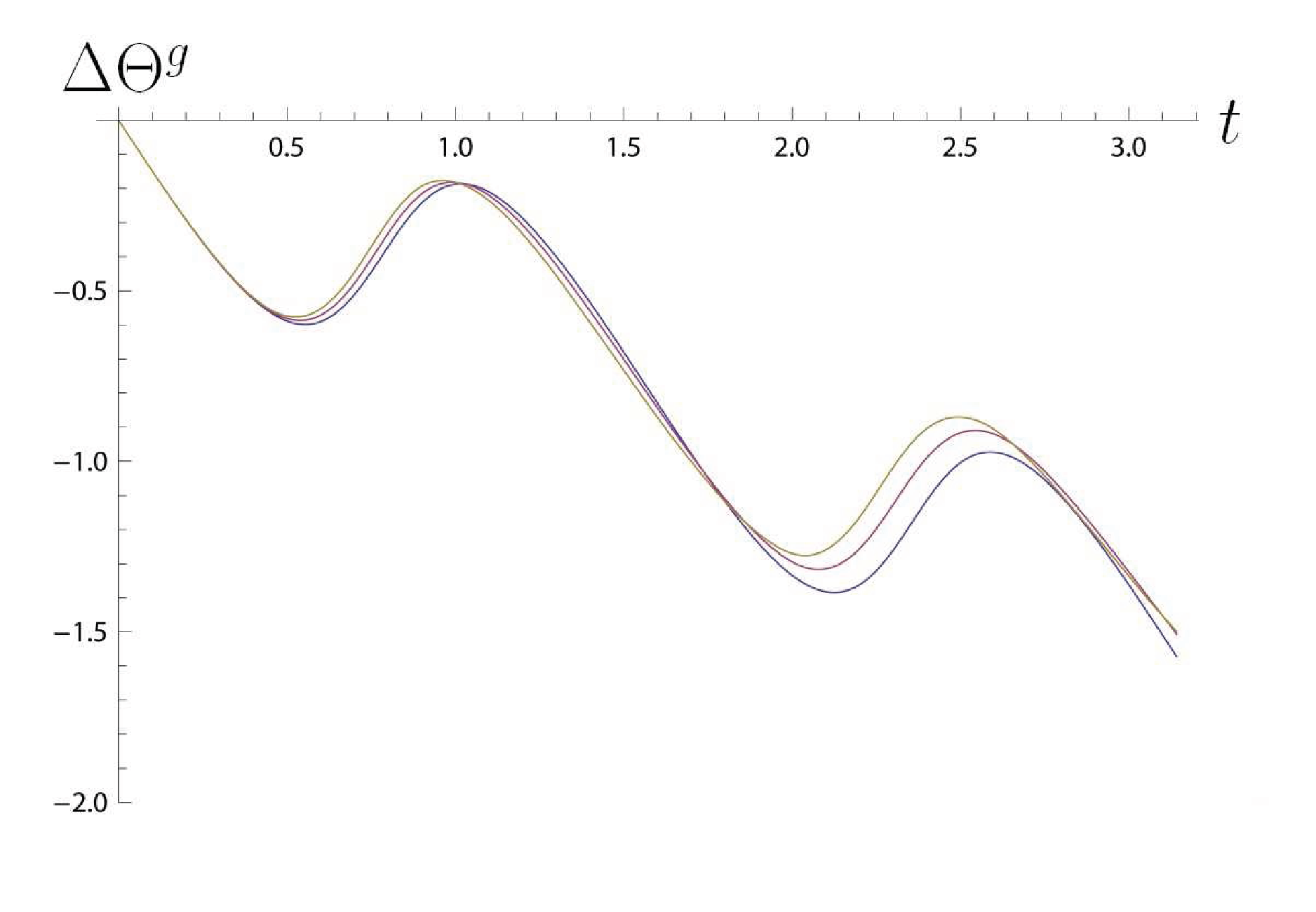}}}
\resizebox*{0.22\textheight}{!}{
{\includegraphics{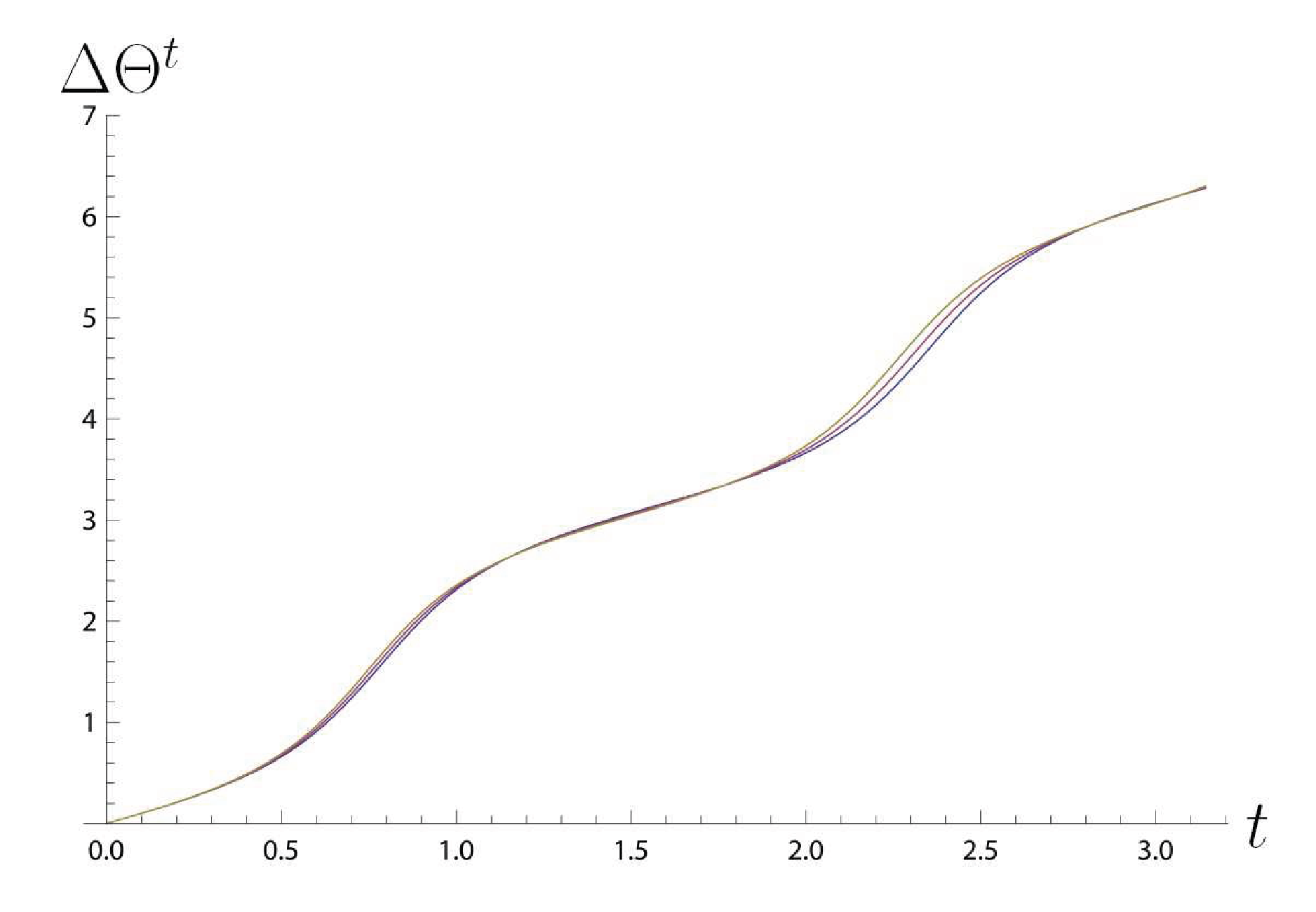}}}
\caption{\textsl{(Color online) The EL invariants in the range $t\in[0,\pi]$ for the harmonic oscillator with $\Omega_0=2$ (top left) the dynamic phase (top right), geometric phase (bottom left), and total phase (bottom right).
In blue, the corresponding graphics for a noiseless oscillator. The magenta colour corresponds to the $m=1$ multiplicative noise of amplitude $\alpha_\Omega=0.1$ and the green colour corresponds to the amplitude $\alpha_\Omega=0.2$.}}
\label{Set1a}
\end{center}
\end{figure}

We choose first the particular case of the pure harmonic oscillator with $\Omega_0=2$ and present
the effect of the multiplicative noise on the EL invariant and the three phases
in Fig.~(\ref{Set1a}). 
One can see that the general effect is a small shift proportional with the amplitude of the noise. For $\rho$ we use the following formula obtained by Eliezer and Gray using an auxiliary plane motion \cite{Eliezer}
\begin{equation}\label{elgray1}
\rho=\sqrt{x^2(t)+\frac{h^2}{\alpha^2}x_2^2(t)}~,
\end{equation}
where $h$ is an arbitrary constant playing the role of the constant angular momentum and
\begin{equation}\label{elgray2}
x(t)=\alpha x_1(t)+\beta x_2(t)
\end{equation}
is the general harmonic solution which satisfy the initial conditions $x(0)=\alpha$ and $\dot{x}(0)=\beta$. In this paper, we choose $\alpha=1$ and $\beta=0$ as initial conditions and fix the constant $h$ to unity, which means that we use a quadrature formula for $\rho$
\begin{equation}\label{elgray3}
\rho(t)=\sqrt{x_1^2(t)+x_2^2(t)}~.
\end{equation}
The corresponding harmonic solutions are $x_1=\cos \Omega_0 t$ and $x_2=\frac{1}{\Omega_0}\sin \Omega_0 t$ and it can be easily shown that the Milne-Pinney function of the pure harmonic oscillator which fulfills the initial conditions $\rho(0)=1$ and $\dot{\rho}(0)=0$ is
\begin{equation}\label{rhoho}
\rho(t)=\frac{\sqrt{(\Omega_0^2-1)\cos^2\Omega_0t+1}}{\Omega_0}~,
\end{equation}
which for $\Omega_0=2$ reduces to $\rho(t)=\frac{1}{2}\sqrt{3\cos^2 2t+1}$.

\begin{figure} [x!] 
\begin{center}
\resizebox*{0.3\textheight}{!}{
{\includegraphics{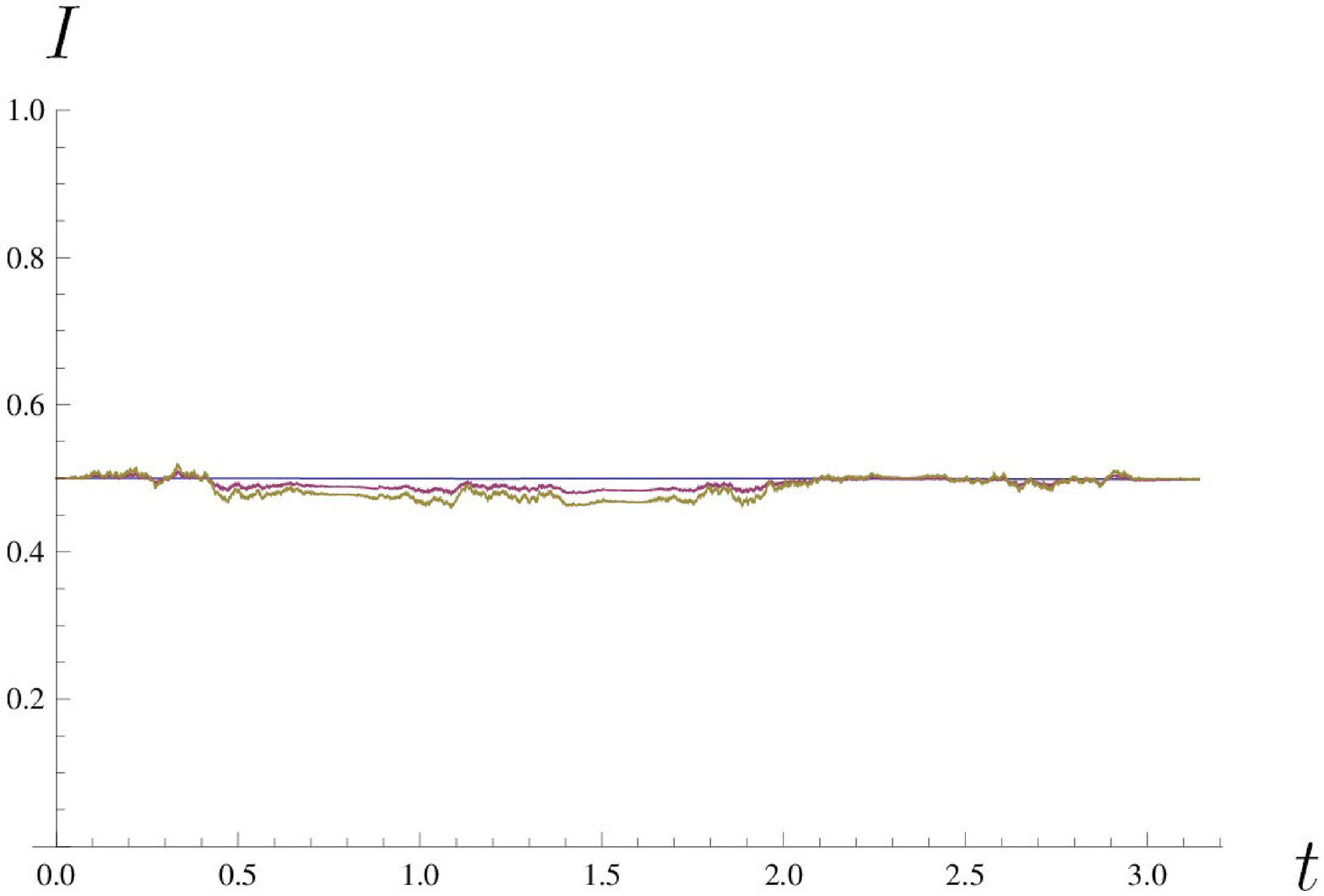}}}
\resizebox*{0.3\textheight}{!}{
{\includegraphics{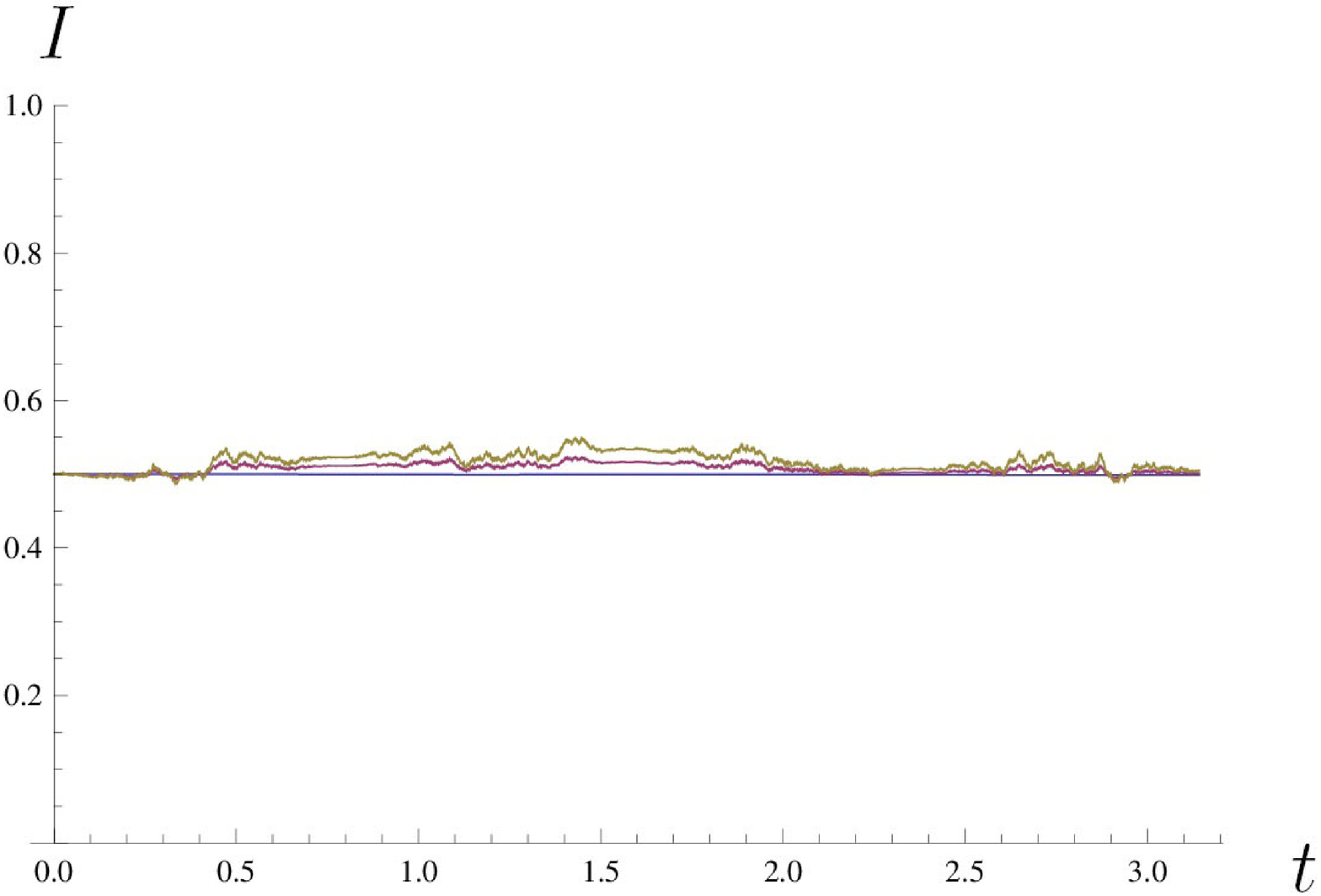}}}
\caption{\textsl{(Color online) The EL invariants in the range $t\in[0,\pi]$ for the harmonic oscillator with $\Omega_0=2$. The noise is included only in the $x$-oscillator (left) and only in the $\rho$-oscillator (right).
The magenta colour corresponds to the $m=1$ multiplicative noise of amplitude $\alpha_\Omega=0.1$ and the green colour corresponds to the amplitude $\alpha_\Omega=0.2$.}}
\label{Set1b}
\end{center}
\end{figure}

For completeness, in Fig.~(\ref{Set1b}) we present the Ermakov-Lewis invariant for the harmonic Ermakov system with the noise included in only one of the oscillators. For these asymmetric cases, we have always obtained a more visible effect of the noise on the Ermakov-Lewis invariant which for us serves as evidence that the correct procedure is to include the noise on equal footing in both equations of the Ermakov systems. 

\renewcommand{\baselinestretch}{1.0}

\renewcommand{\baselinestretch}{1.0}

\subsection{Parametric oscillator with $\Omega(t)=2\sin t$}

\begin{figure}[ht]
\begin{center}
\resizebox*{0.3\textheight}{!}{
{\includegraphics{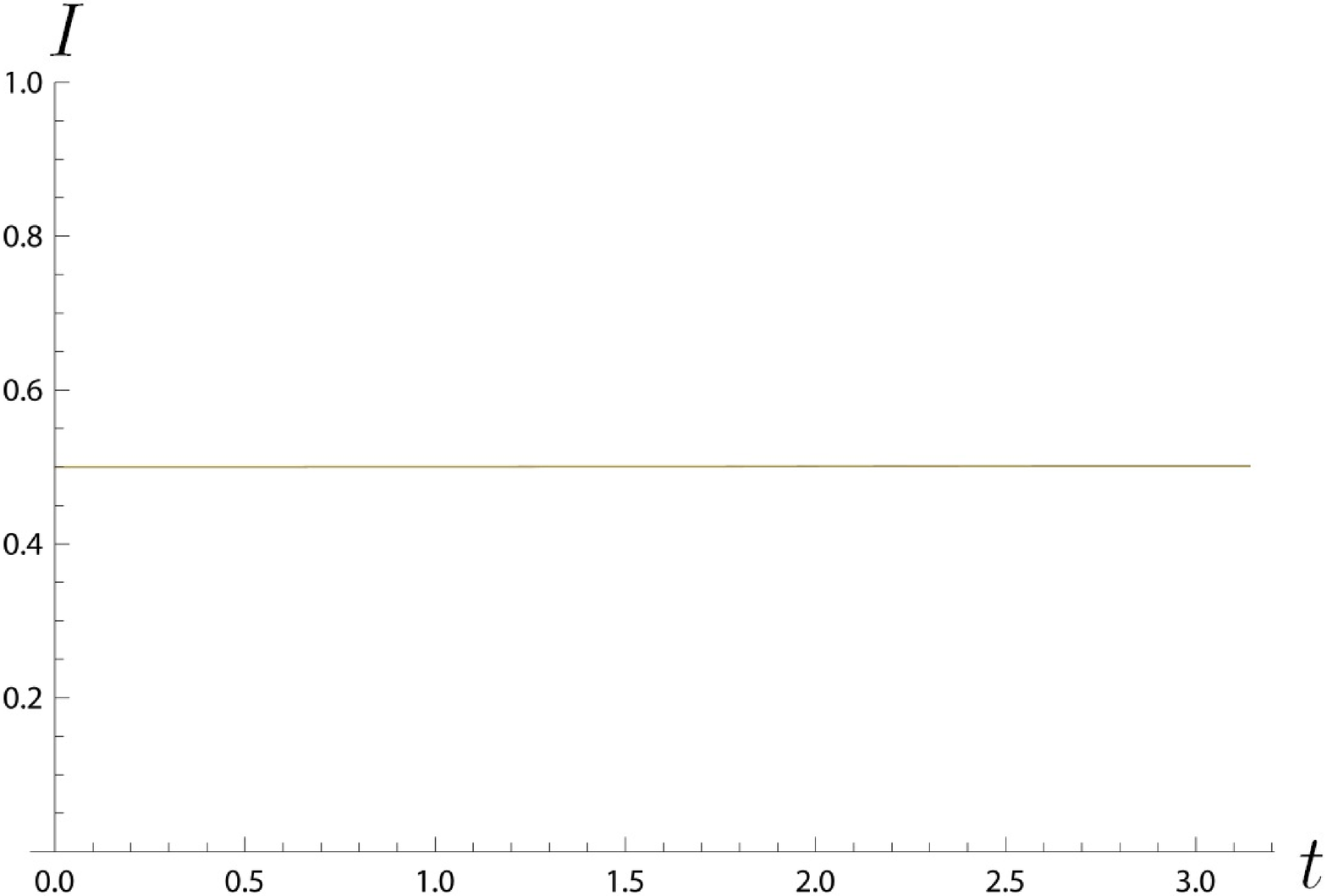}}}
\resizebox*{0.3\textheight}{!}{
{\includegraphics{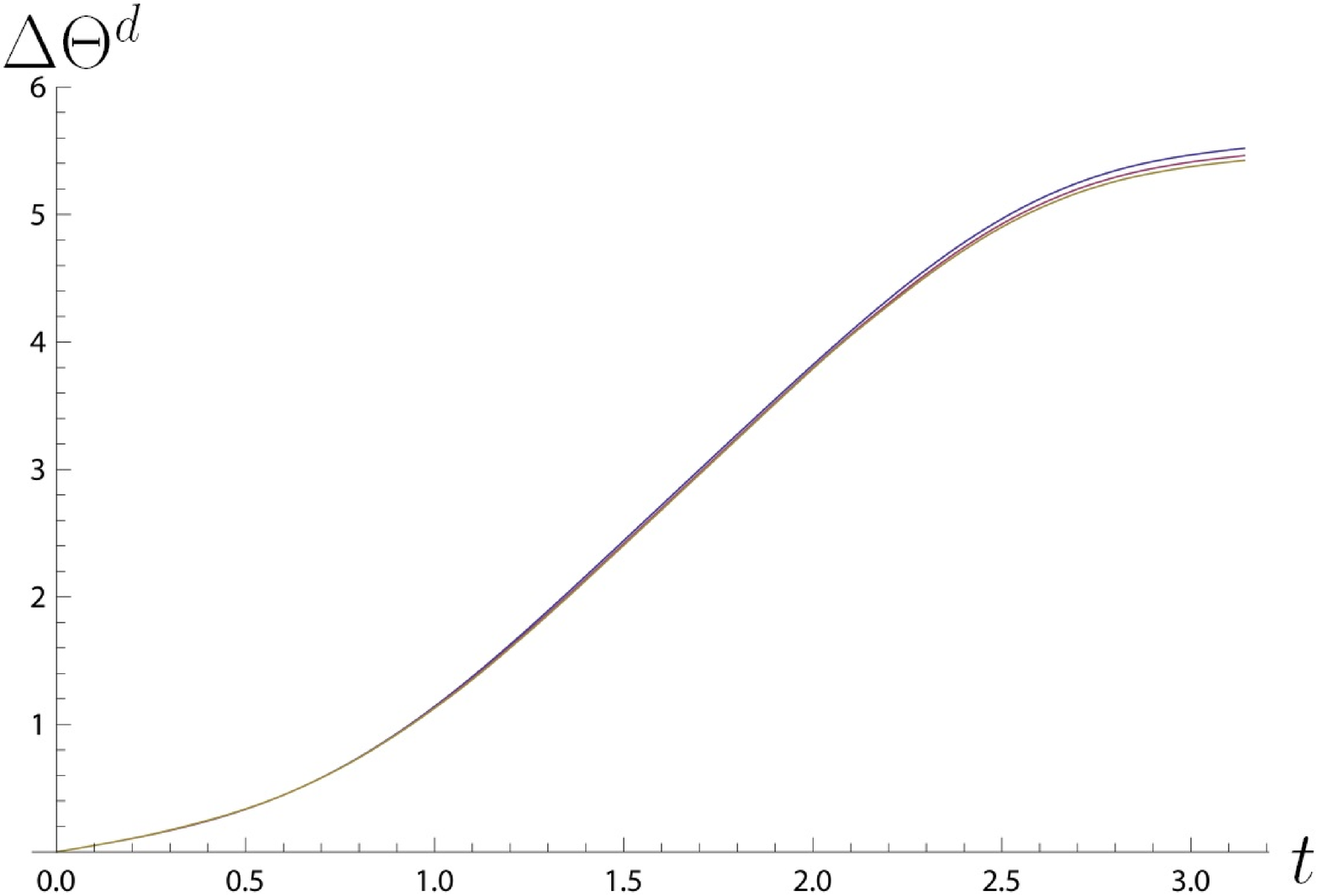}}}
\resizebox*{0.3\textheight}{!}{
{\includegraphics{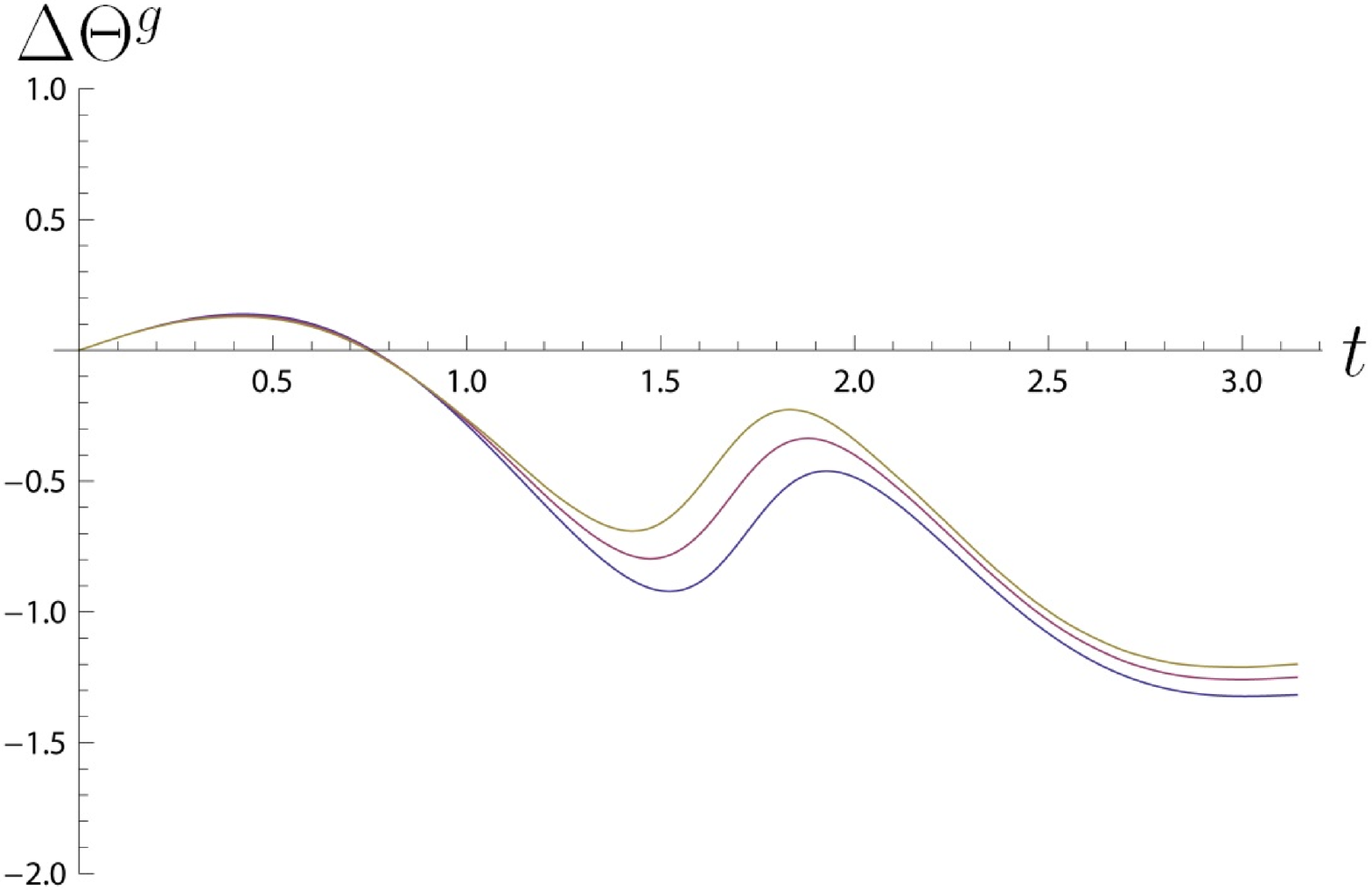}}}
\resizebox*{0.3\textheight}{!}{
{\includegraphics{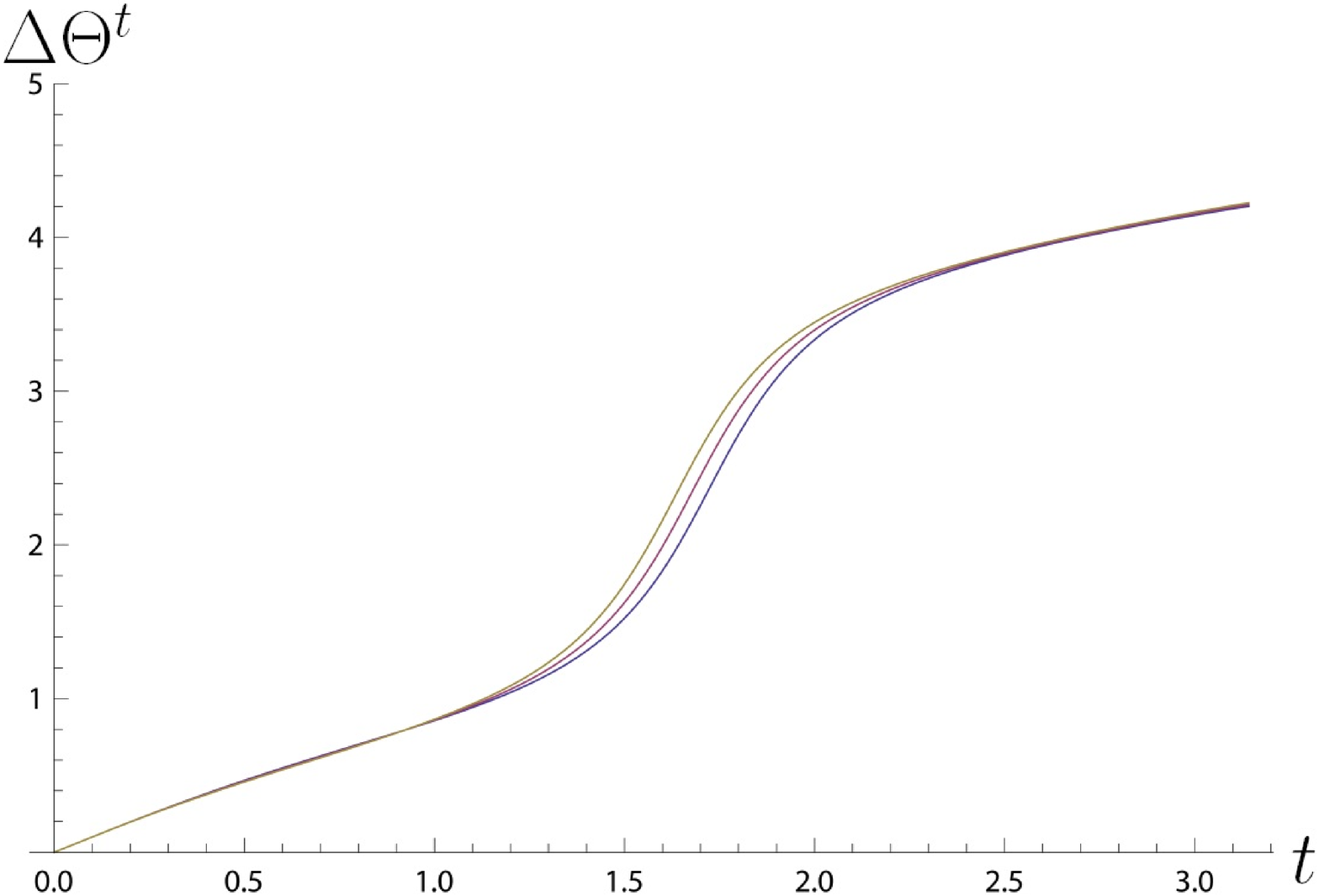}}}
\caption{\textsl{(Color online) The EL invariants $I$ in the range $t\in[0,\pi]$ for a parametric oscillator with $\Omega(t)=2\sin t$
(top left). The dynamic phase (top right), geometric phase (bottom left), and total phase (bottom right) for the noiseless parametric system (blue), $m=1$ multiplicative noise of small amplitude $\alpha_\Omega=0.1$ (magenta), and $\alpha_\Omega=0.2$ (green), respectively.}}
\label{Set2a}
\end{center}
\end{figure}

The plots of the Ermakov quantities for this case are displayed in Fig.~(\ref{Set2a}).
The effect of the noise is similar to the previous case.

For completeness, we also present the analytic solution of this parametric oscillator. The corresponding parametric equation can be easily shown to be of the following Mathieu type
\begin{equation}\label{mathieu1}
\ddot{x}+(2-2\cos2t)x=0~.
\end{equation}
The solutions satisfying the initial conditions as given by Eliezer and Gray are
\begin{equation}\label{mathieu2}
x_1(t)=\frac{C(2,1;t)}{C(2,1;0)}~, \qquad x_2(t)=\frac{S(2,1;t)}{\dot{S}(2,1;0)}~,
\end{equation}
where $C(2,1;0)\approx 1.182$ and $\dot{S}(2,1; 0)\approx 0.5336$ and
the $C$ and $S$ functions are the known Mathieu cosine and Mathieu sine functions and their Wronskian is $W=1$. Therefore, the analytic Milne-Pinney solution fulfilling the same initial conditions as $x_1$ has implicitly the quadrature format $\rho(t)=\sqrt{x_1^2(t)+x_2^2(t)}$.

\renewcommand{\baselinestretch}{1.0}

\subsection{Parametric oscillator with $\Omega(t)=2t^2$}

\begin{figure}[ht]
\begin{center}
\resizebox*{0.3\textheight}{!}{
{\includegraphics{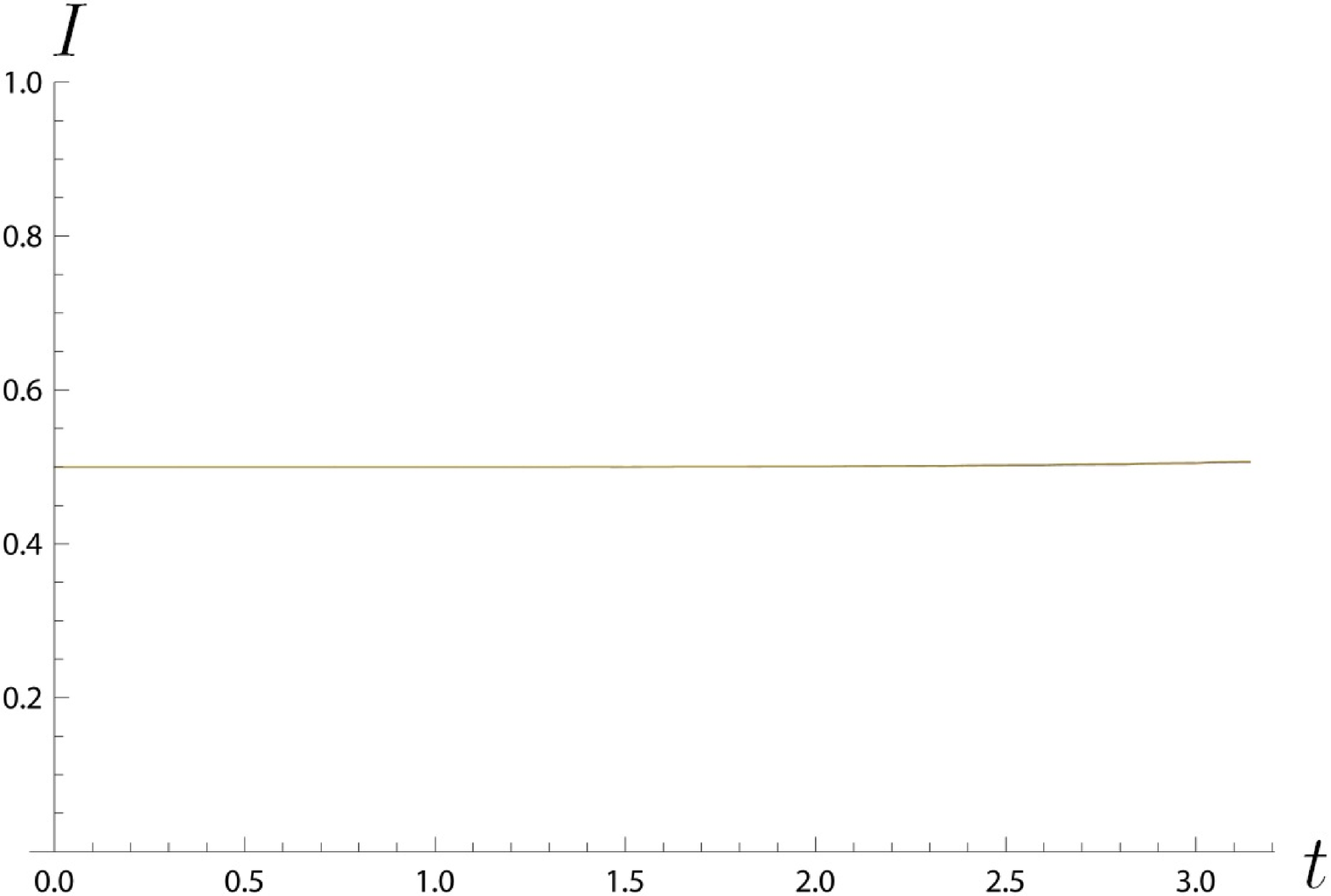}}}
\resizebox*{0.3\textheight}{!}{
{\includegraphics{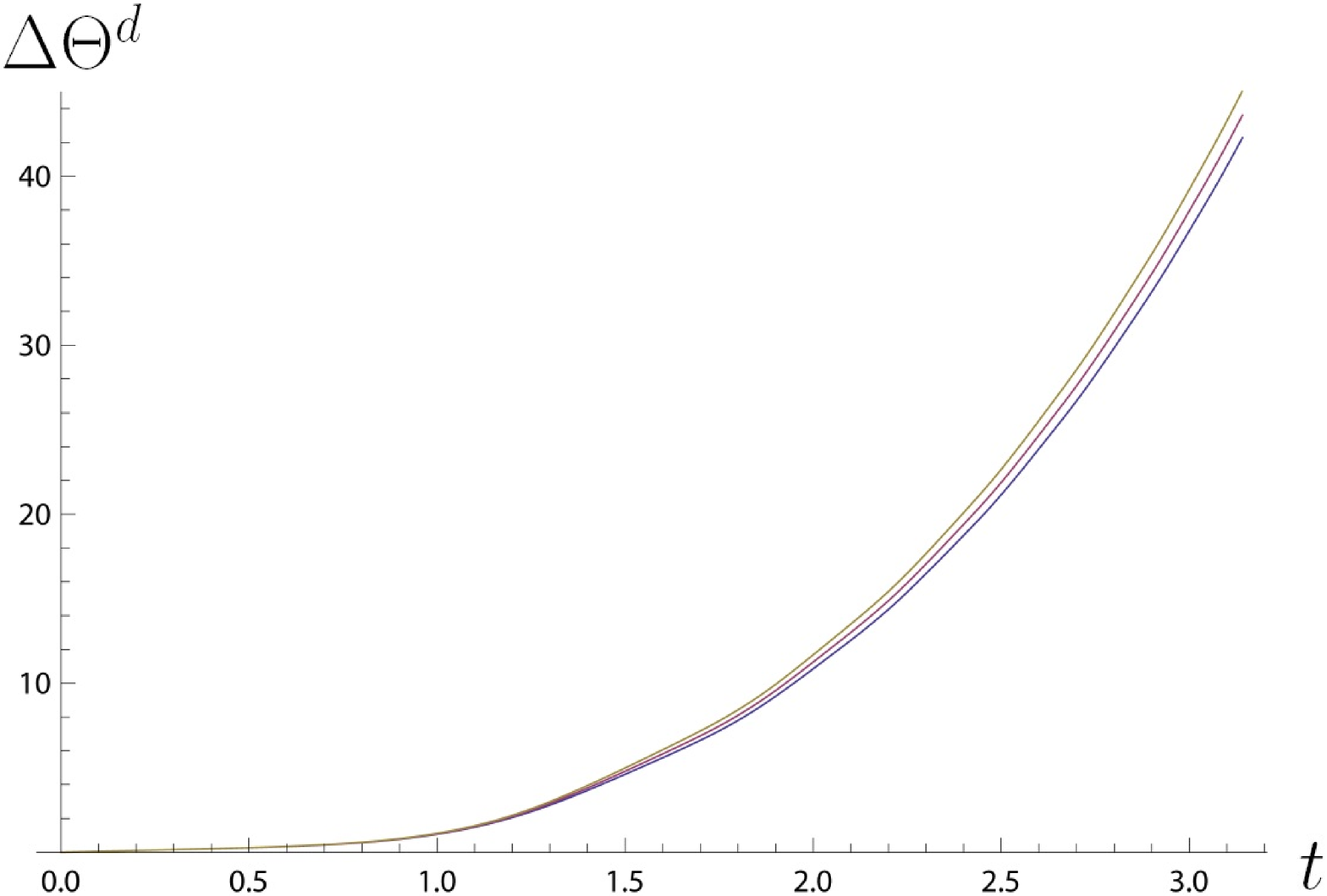}}}
\resizebox*{0.3\textheight}{!}{
{\includegraphics{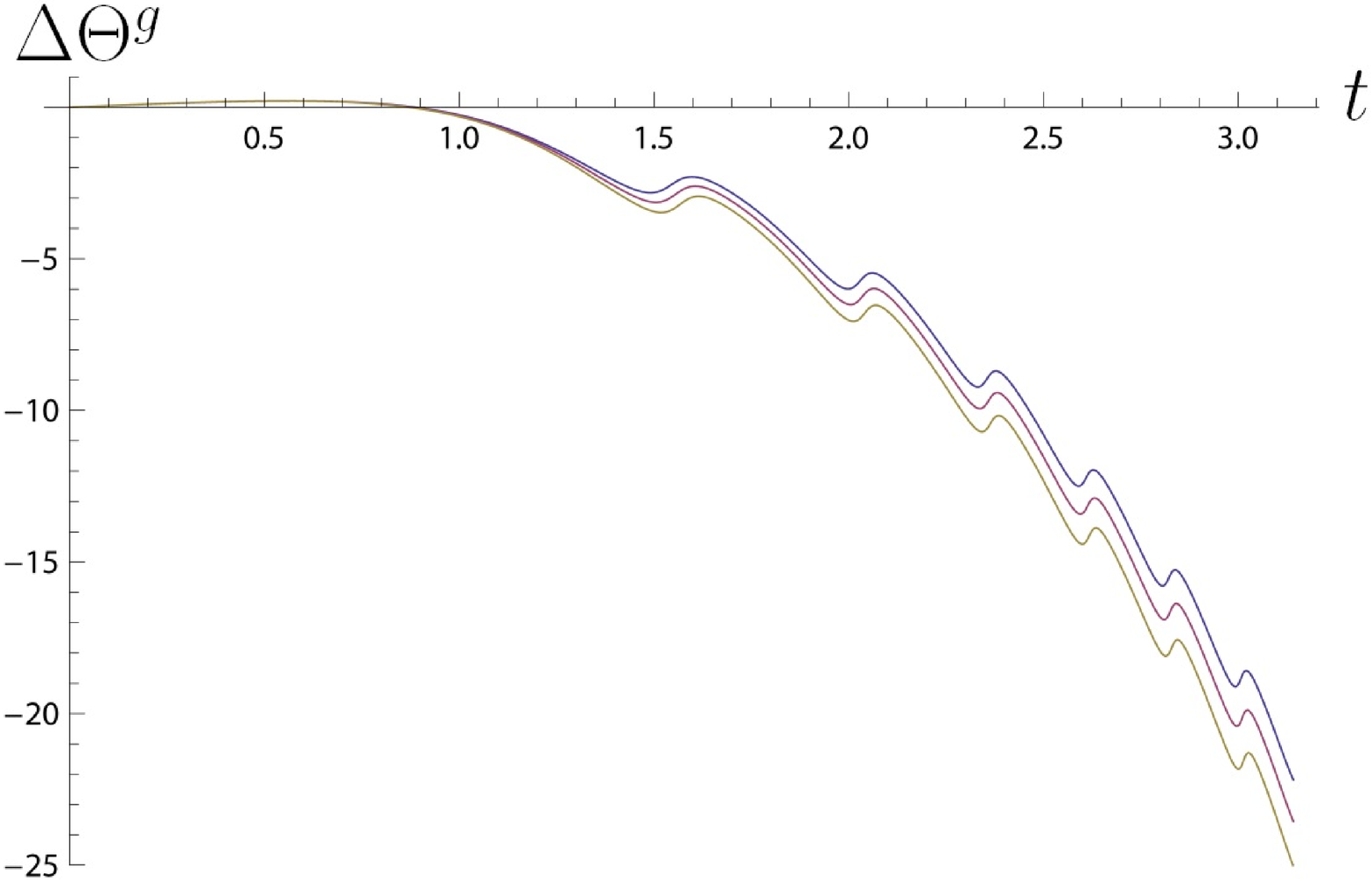}}}
\resizebox*{0.3\textheight}{!}{
{\includegraphics{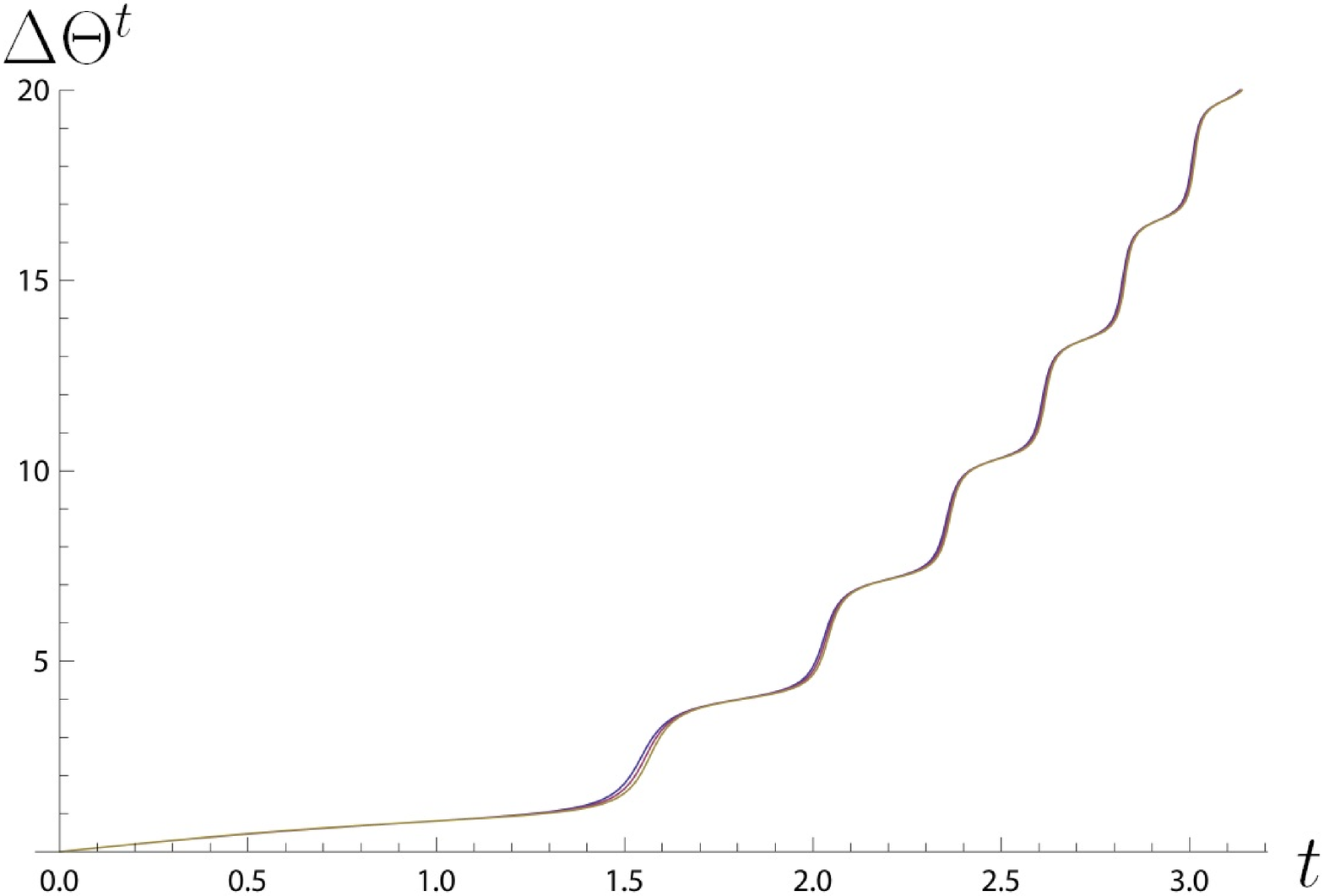}}}
\caption{\textsl{(Color online) The EL invariant $I$ in the range $t\in[0,\pi]$ for the parametric oscillator with $\Omega(t)=2t^2$ (top left).
The dynamic phase (top right), geometric phase (bottom left), and total phase (bottom right) for the noiseless case (blue), $m=1$ noise of amplitudes $\alpha_\Omega=0.1$ (magenta) and $\alpha_\Omega=0.2$ (green), respectively.}}
\label{Set3a}
\end{center}
\end{figure}

This case can be found in the list of analytic cases given by Eliezer and Gray in their paper. We provide here the two linearly independent solutions of the parametric equation which satisfy the Eliezer-Gray initial conditions:
\begin{equation}\label{p1}
x_1(t)=\frac{\Gamma\left(\frac{5}{6}\right)}{3^{\frac{1}{6}}}\sqrt{t}J_{-\frac{1}{6}}\left(\frac{2t^3}{3}\right)~, \qquad x_2(t)=\frac{\Gamma\left(\frac{1}{6}\right)}{2\cdot 3^{\frac{5}{6}}}\sqrt{t}J_{\frac{1}{6}}\left(\frac{2t^3}{3}\right)~.
\end{equation}
Their Wronskian is $W=1$ and therefore one can obtain the Milne-Pinney function from an equation of the same form as (\ref{elgray3}).
The plots of the Ermakov quantities for this case are presented in Fig.~(\ref{Set3a}).
The slight shift effect of the noise is again noticeable.

\renewcommand{\baselinestretch}{1.0}

\section{Concluding remarks}\label{Sconc}

We have studied the effects of the simplest type of multiplicative noise on the Ermakov systems for three particular cases.
The Euler-Maruyama numerical scheme has been used to include the stochastic noise in the Ermakov system.
It has been found that the usage of the same noise term in the two oscillators of the Ermakov system leads to a cancelation of the noise effects on the Ermakov-Lewis invariant, which can be due to the structure of the invariant itself. On the other hand, the effect of the noise in the nonlinear oscillator leads to a shift effect on the three phases of the system.
This can be understood as a consequence of the averaging effect produced by the integrals in the expressions of the phases.

 \subsection*{Acknowledgements}
The authors wish to thank one of the referees whose comments led to a substantial improvement of this work.

\bigskip
\bigskip


\end{document}